\documentclass{webofc}
\usepackage[varg]{txfonts}   % Web of Conferences font

\newcommand{\mathsym}[1]{{}}
\newcommand{\unicode}[1]{{}}

\newcommand{\beq}{\begin{equation}}
\newcommand{\eeq}{\end{equation}}
\newcommand{\bea}{\begin{eqnarray}}
\newcommand{\eea}{\end{eqnarray}}

\begin{document}
\title{Non-perturbative Approach to Equation of State and Collective Modes of the QGP}

\author{\firstname{Shuai Y.F.} \lastname{Liu}\inst{1}\fnsep\thanks{\email{lshphy@gmail.com}} \and
        \firstname{Ralf} \lastname{Rapp}\inst{1}\fnsep\thanks{\email{rapp@comp.tamu.edu}} %\and
%        \firstname{Third author} \lastname{Third author}\inst{3}\fnsep\thanks{\email{Mail address for last
  %           author if necessary}}
        % etc.
}

\institute{Cyclotron Institute and Department of Physics \& Astronomy, Texas A\&M University, College Station, TX 77843-3366, USA}

\abstract{We discuss a non-perturbative $T$-matrix approach to investigate the microscopic structure 
of the quark-gluon plasma (QGP).  Utilizing an effective Hamiltonian which includes both light- and 
heavy-parton degrees of freedoms. The basic two-body interaction includes color-Coulomb and confining 
contributions in all available color channels, and is constrained by lattice-QCD data for the heavy-quark 
free energy. The in-medium $T$-matrices and parton spectral functions are computed selfconsistently 
with full account of off-shell properties encoded in large scattering widths. We apply the $T$-matrices 
to calculate the equation of state (EoS) for the QGP, including a ladder resummation of the Luttinger-Ward functional using a matrix-log technique to account for the dynamical formation of bound states. 
It turns out that the latter become the dominant degrees of freedom in the EoS at low QGP temperatures 
indicating a transition from parton to hadron degrees of freedom. The calculated spectral properties 
of one- and two-body states confirm this picture, where large parton scattering rates dissolve the 
parton quasiparticle structures while broad resonances start to form as the pseudocritical temperature 
is approached from above. Further calculations of  transport coefficients reveal a small viscosity and 
heavy-quark diffusion coefficient. 
}
\maketitle
\section{Introduction}
\label{intro}
\begin{figure}[!t]
\begin{center}
\includegraphics[width=0.6\columnwidth]{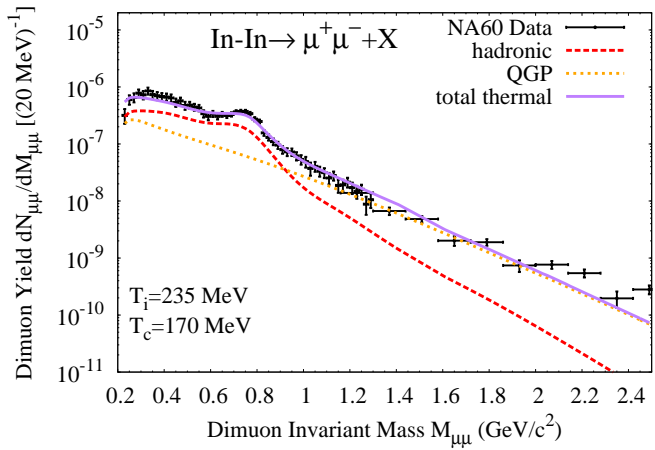}
\caption
{Dimuon excess spectra (after subtraction of background and the final-state hadron decay
cocktail)~\cite{Specht:2010xu} compared to theoretical calculations of thermal radiation from an 
expanding fireball with contributions from the QGP and hadronic phase with in-medium $\rho$ and 
$\omega$ spectral functions; figure taken from Ref.~\cite{Rapp:2014hha}.}
\label{fig_na60}
\end{center}
\end{figure}

The theoretical study of the QCD phase diagram poses formidable challenges due to the strong force 
between quarks and gluons at intermediate and large distances. The QCD phase structure is believed 
to be tightly connected to two key phenomena of the standard model, namely hadronic mass generation 
and color confinement. In addition, it turned out that the strongly interacting fireball medium 
formed in ultra relativistic collisions of heavy nuclei possesses the smallest known ratio of 
viscosity to entropy density, giving rise to the notion of the strongly coupled quark-gluon 
plasma (sQGP)~\cite{Shuryak:2014zxa,Braun-Munzinger:2015hba,Heinz:2015lpa}, a near-perfect liquid. 
The microscopic structure of the sQGP, including its prevalent degrees of freedom and its possible 
relation to the nearby phase transition(s) into hadronic matter, remains a forefront topic in 
contemporary research.
In addition to the low viscosity inferred from fluid-dynamic models for the bulk evolution of the 
medium in heavy-ion collisions at RHIC and the LHC, the large modifications observed for the spectra 
and elliptic flow of heavy-flavor hadrons~\cite{Prino:2016cni} have provided a more direct evidence 
of a frequent rescattering that (heavy) quarks undergo throughout the expansion of the nuclear 
fireball. At the same time, dilepton invariant-mass spectra emanating from the thermal radiation
of the hot system~\cite{Arnaldi:2006jq,Tserruya:2009zt,Specht:2010xu} have revealed that the
$\rho$-meson resonance peak, which dominates in the low-mass part of the electromagnetic spectral 
function in vacuum, melts due to strong rescattering in hot and dense hadronic 
matter~\cite{Rapp:2009yu}, cf.~Fig.~\ref{fig_na60}. 
While this melting provides a signature of chiral symmetry restoration~\cite{Hohler:2013eba}, it
is also suggestive for a transition in the degrees of freedom in the medium, from well-defined
hadronic states into a quasi-continuum of partons.
The challenge remains how this melting can be seamlessly connected to a quark-gluon based 
description on the QGP side of the medium.  

%The perturbative QCD estimation for the interaction strength is not sufficient to explain these 
%unique features found in the experiments, 
A microscopic description of the sQGP is generally expected to require nonperturbative ingredients,
in terms of both the underlying interaction (beyond Color-Coulomb) and resummations of diagrams.
The vicinity of the sQGP to the phase transition into hadrons suggests that (remnants of) a 
confining force and the emergence of bound states play an essential role. Computations of the 
heavy-quark (HQ) free energy in lattice QCD show that its long-distance limit stays above zero
until temperatures of about 450\,MeV~\cite{Bazavov:2014pvz}, supporting to the presence of 
non-Coulombic contributions (the leading Coulombic contribution is negative).  
To capture these aspects, we have developed a thermodynamic $T$-matrix approach which resums the
ladder series of an in-medium interaction kernel that includes the screened Coulomb and confining 
forces to characterize the interaction strength between the partons in the QGP, and allows for
the dynamical formation of bound and resonance states. It provides a uniform treatment of light and
heavy partons, can be constrained by pertinent results from lQCD and enables the calculation
of spectral functions and transport coefficients for applications to heavy-ion phenomenology.   
We will briefly review the selfconsistent $T$-matrix approach in Sec.~\ref{sec_Tmat}, discuss 
the lattice-QCD (lQCD) constraints on the underlying potential in Sec.~\ref{sec_pot}, compute 
the equation of state (EoS) in Sec.~\ref{sec_eos}, and evaluate the emerging spectral and transport 
properties in Secs.~\ref{sec_spec} and \ref{sec_trans}, respectively. Section~\ref{sec_concl} 
contains our conclusions and outlook. 

%%%%%%%%%%%%%%%%%%%%%%%%%%%%%%%%%%%%%%%%%%
\section{$T$-matrix approach}
\label{sec_Tmat}
\begin{figure}[!t]
    \begin{center}
\includegraphics[width=0.75\columnwidth]{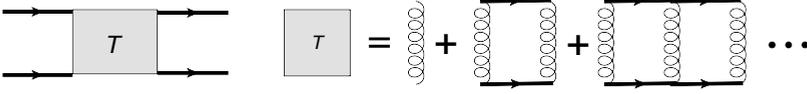}
\caption{\(T\)-matrix resummation of ladder diagrams.}
\label{fig_Tmatrix}
        \end{center}
\end{figure}

%%%%%%%%%%%%%%%%%%%%%%%%%%%%%%%%%%%%%%%%%
We start from an effective in-medium parton Hamiltonian, 
\begin{align}
&H=\sum\varepsilon(\textbf{p})\psi^\dagger(\textbf{p})\psi (\textbf{p})+
\frac{1}{2}\psi^\dagger(\frac{\textbf{P}}{2}-\textbf{p})\psi^\dagger(\frac{\textbf{P}}{2}+\textbf{p})
V \psi(\frac{\textbf{P}}{2}+\textbf{p}')\psi(\frac{\textbf{P}}{2}-\textbf{p}') \ , 
\label{eq_Hqgp}               
\end{align}
with single-particle energies $\varepsilon(\textbf{p})=\sqrt{M^2+\textbf{p}^{2} }$ containing the
bare parton masses, $M_i$  and a 2-body interaction kernel, $V$, where $\textbf{p}, \textbf{p}'$ 
and $\textbf{P}$ denote the relative and total momenta of the pair.
The sum is over momentum, spin, color, and particle species
(3 light-quark flavors, gluons, charm and bottom quarks). Summing up the ladder diagrams generated 
from this Hamiltonian, we obtain the $ T $-matrix 
equation~\cite{Mannarelli:2005pz,Riek:2010fk,Riek:2010py},   
\begin{align}
&T(z,\textbf{p},\textbf{p}')=V(\textbf{p},\textbf{p}')+
\int_{-\infty}^{\infty}\frac{d^3\textbf{k}}{(2\pi)^3}V(\textbf{p},\textbf{k})
G^{0}_{(2)}(z,\textbf{k})T(z,\textbf{k},\textbf{p}') 
\label{eq_T}
\end{align}
in the center-of-mass (CM) frame ($\textbf{P}$=0) with external two-body energy 
$z$, cf.~Fig.~\ref{fig_Tmatrix}.  
In the Matsubara representation, 
the one- and uncorrelated two-body propagator take the form 
\begin{equation}
G(i\omega_n,\textbf{k})=\frac{1}{[G^{0}(i\omega_n,\textbf{k})]^{-1}-\Sigma(i\omega_n,\textbf{k})} \ , \ 
G^{0}_{(2)}(iE_n,\textbf{k})=-\beta^{-1}\sum_{\omega_n} G(iE_n-i\omega_n,\textbf{k})G(i\omega_n,\textbf{k}) 
\ , 
\label{eq_G12} 
\end{equation}
respectively, and $ G^{0}=1/(i\omega_n-\varepsilon(\textbf{k}))$ is the bare propagator.
The selfenergy is obtained by closing $T$-matrix, 
\begin{equation}
\Sigma(iw_{n})=\int d\tilde{p}~\,T G\equiv~
-\beta^{-1}\sum_{\nu_n}\int \frac{d^{3}\textbf{p}}{(2\pi)^3}T(i\omega_{n}+i\nu_{n})G(i\nu_n) \ ,  
\label{eq_selfE} 
\end{equation}
which we evaluate using standard spectral representations. 
Equations (\ref{eq_T}) and (\ref{eq_selfE}) form a selfconsistency problem that we solve
by numerical iteration, thereby satisfying thermodynamic conservation laws~\cite{Baym:1961zz}. 

The key {\em inputs} to the Hamiltonian are the 2-body interaction kernel, $V$,  and the bare 
parton masses, $M$. In the following two sections we discuss how we constrain them using 
lQCD data for the HQ free energy and the EoS.

%%%%%%%%%%%%%%%%%%%%%%%%%%%%%%%%%%%%%%%%%%
\section{Interaction kernel and HQ free energy}
\label{sec_pot}
%%%%%%%%%%%%%%%%%%%%%%%%%%%%%%%%%%%%%%%%
\begin{figure}[!t]
        \begin{center}
                \includegraphics[width=0.67\columnwidth]{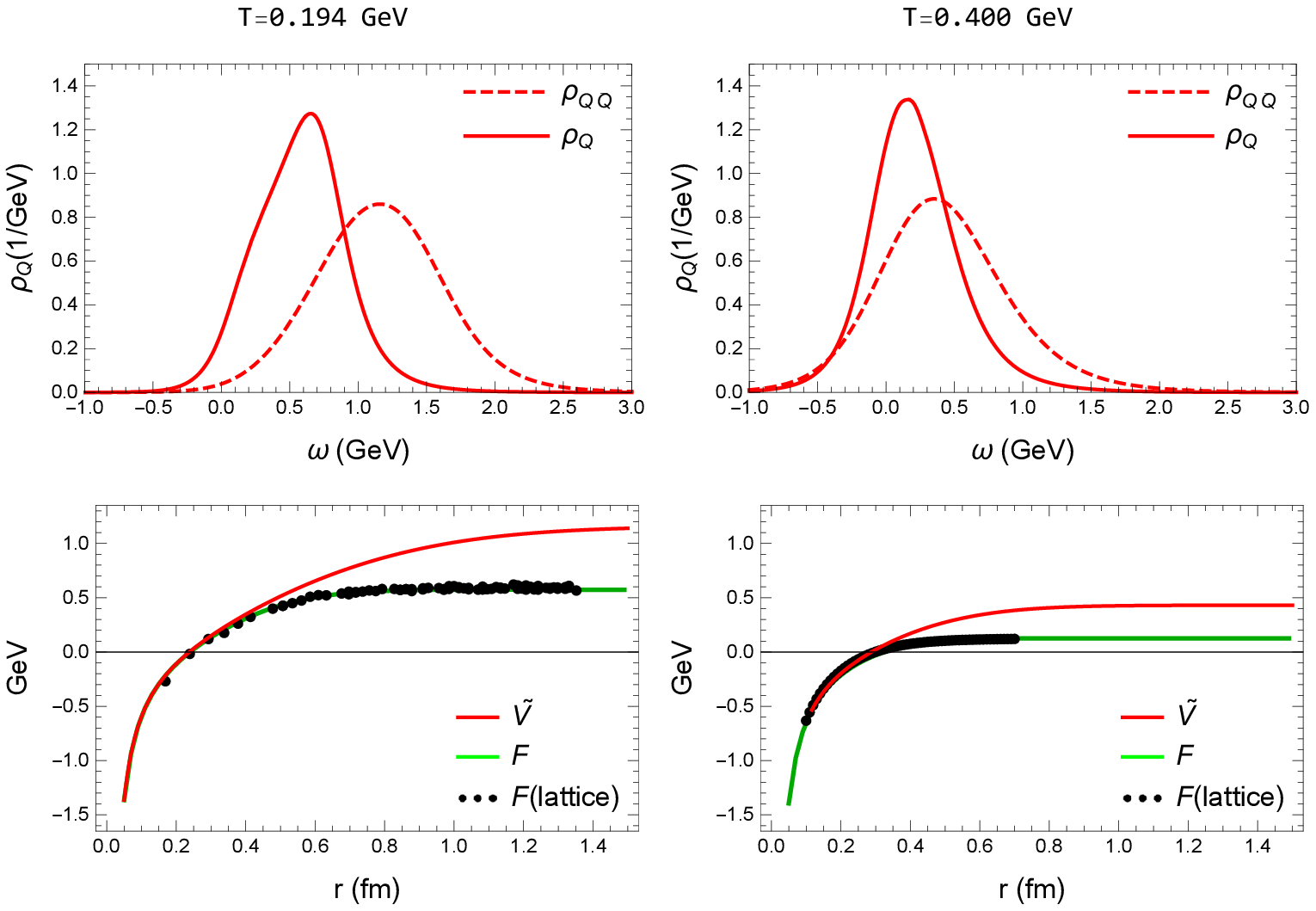}
                \includegraphics[width=0.31\columnwidth]{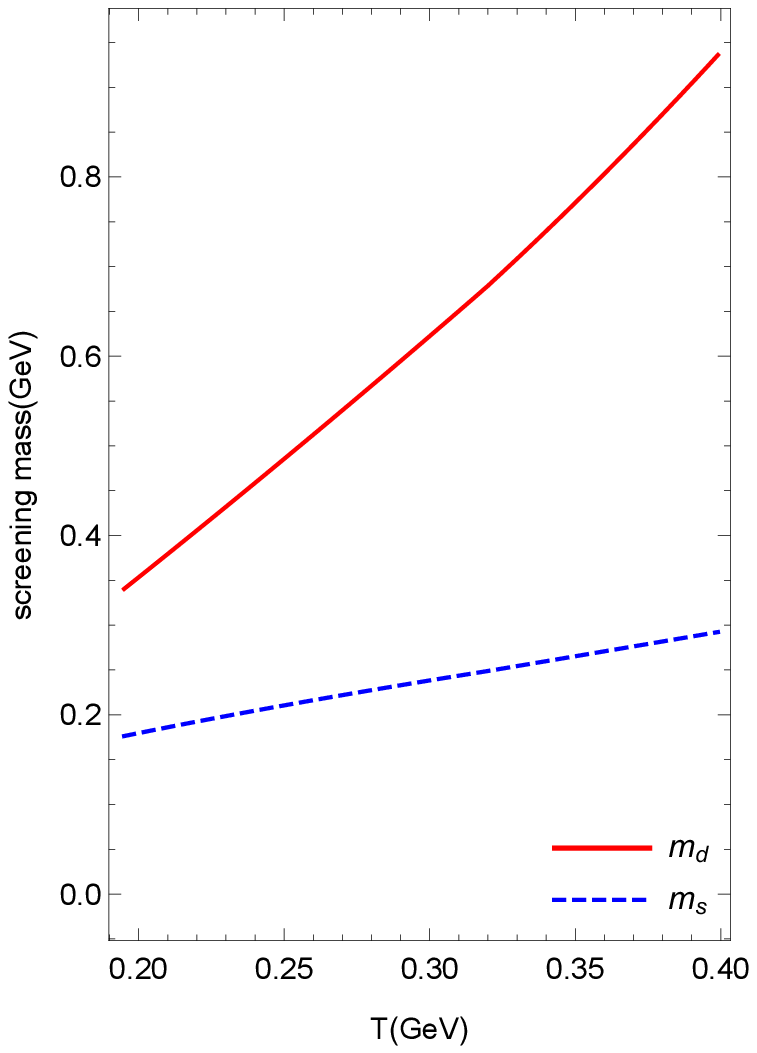}
                \caption{Our results for the selfconsistent fit to lQCD data~\cite{Mocsy:2013syh} 
for the static HQ free energy and resulting potential, $\tilde{V}(r)$ (lower left and middle 
panels for two different temperatures) and the associated one- and uncorrelated two-body spectral 
functions, $\rho_{Q}(\omega,\infty)$ and $\rho_{Q\bar Q}(\omega,\infty)$, respectively (upper 
left and middle panels for the same two temperatures), for the strongly-coupled solution. The 
right column shows the fitted screening masses for Coulomb (solid line) and string (dashed line) 
interactions.}
                \label{fig_scs-V-F}
        \end{center}
\end{figure}
In Ref.~\cite{Liu:2015ypa} we have obtained a relation between the static HQ free energy and 
the potential within the $T$-matrix formalism, given by
\begin{align}
&F_{Q\bar{Q}}(r,\beta)=\frac{-1}{\beta}\ln \bigg[\int_{-\infty}^{\infty} 
dE\,e^{-\beta E} \frac{-1}{\pi}\text{Im}[\frac{1}{E+i\epsilon-\tilde{V}(r)
-\Sigma_{Q\bar Q}(E+i\epsilon,r)}]\bigg] \ .
\label{eq_FreeEfinal}               
\end{align}
In the weak-coupling limit (or at low temperatures), where the imaginary part of the in-medium 
two-body selfenergy, $\Sigma_{Q\bar Q}$, is small, one readily recovers that the free energy 
coincides with
the potential, $\tilde{V}$ (defined as containing a non-trivial infinite-distance limit, 
see Eq.~(\ref{Vtilde}) below). However, at strong coupling large imaginary part develop
and lead to appreciable deviations between potential and free energy.   
As an ansatz for the potential, we employ a generalized in-medium (screened) Cornell 
potential~\cite{Megias:2007pq}, 
\begin{equation}
\tilde V(r)=V_\mathcal{C}(r)+V_\mathcal{S}(r)+2\Delta M_Q=-\frac{4}{3}\alpha_s \frac{e^{-m_d r}}{r}
-\frac{\sigma e^{-m_s r- (c_b m_s r)^2}}{m_s}-\frac{4}{3}\alpha_s m_d +\sigma m_s,
\label{Vtilde}
\end{equation}
including color-Coulomb ($V_\mathcal{C} $) and string terms ($V_\mathcal{S}$). Equation (\ref{Vtilde})
is written in the color-singlet channel for static quarks, but in the calculations of the EoS
reported in  the next section we include all possible two-body color channels with
appropriate Casimir factors and relativistic corrections~\cite{Liu:2017qah}.  The screening of 
the string term is constrained using $ m_s= (c_s m_d^2 \sigma/\alpha_s)^{1/4} $~\cite{Liu:2016ysz}
and also contains a quadratic term, $  (c_b m_s r)^2 $, to better mimic string breaking effects. 
The genuine two-body part is defined as $V(r)=V_\mathcal{C}(r)+V_\mathcal{S}(r)$, while
the nonzero  infinite-distance limit, $\tilde V(\infty)$, is related 
with the mass generated by self-interactions, 
\begin{equation}
\Delta M_Q=-\frac{1}{2}\int dr \rho(r)V(r) =\frac{1}{2}(-\frac{4}{3}\alpha_s m_d +\sigma m_s)
=\tilde V(\infty)/2 \ , 
\label{eq_fockmass}
\end{equation}
corresponding to the classical static potential energy of a point charge, \(\rho(r)=\delta(r)\), 
in its own potential; \(\Sigma_{Q\bar Q}(z,r)\) denotes the analytical two-body selfenergy 
whose $r$-dependence includes interference effects~\cite{Laine:2006ns,Liu:2017qah}  which 
ensure that \(\Sigma_{Q\bar Q}(z,r)\) vanishes for a color-singlet $Q\bar Q$ pair in the 
$r\to0$ limit. 

The solution for a fit of the underlying potential to the lQCD free-energy data turns
out to not be unique. We therefore bracket the possible range of potentials by a 
``strongly-coupled solution" (SCS) and which maximally deviates from the free energy, and 
a ``weakly-coupled solution" (WCS) which is as close as possible to the free energy (within
our ability to find fits). In Fig.~\ref{fig_scs-V-F} we show the results for the fit to the free
energy as well as the underlying potential and one- and uncorrelated two-body spectral 
functions of static quarks for the SCS. The potential is seen to significantly exceed the free 
energy at intermediate and large distances (especially at low temperatures), due to large imaginary 
parts of the selfenergies which in turn lead to broad the spectral functions.  
The latter are characterized by 1- and 2-body widths of near 1 and 2\,GeV,
respectively, which, remarkably, decrease from the lower to the higher temperature. While
the screening mass for the Coulomb term increases quite strongly with temperature (roughly in 
accord with perturbative estimates, $m_d\sim gT$), the screening of the string term is rather
weak. This leads to a long-range force which enables the static quark to interact with multiple 
thermal partons and thus significantly contributes to the large imaginary parts.  
In the WCS (not shown), the potential is much closer to the free energy and the imaginary parts 
are much smaller at low temperatures.

%%%%%%%%%%%%%%%%%%%%%%%%%%%%%%%%%%%%%%%5
\section{Equation of state and light-parton masses}
\label{sec_eos}
\begin{figure}[!t]
	\begin{center}
		\includegraphics[width=0.55\columnwidth]{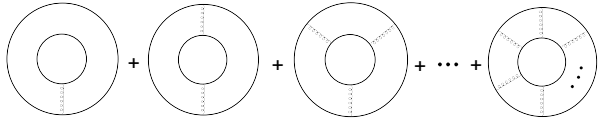}
	\caption{Examples of diagrams that are resummed 
	by the generalized \(T\)-matrix approach for the EoS.}
		\label{fig_wheel}
	\end{center}
\end{figure}

%%%%%%%%%%%%%%%%%%%%%%%%%%%%%%%%%%%%%%%%
Next we apply the Hamiltonian to calculate the EoS of the QGP using the Luttinger-Ward-Baym 
formalism~\cite{PhysRev.118.1417,Baym:1961zz,Baym:1962sx} where the free energy of the system 
is given by
\begin{equation}
\Omega = \mp\sum\text{Tr}\{\ln(-G^{-1})+[(G^0)^{-1}-G^{-1}] G\}\pm\Phi \ , 
\label{Omega}
\end{equation}
where the upper (lower) signs corresponds to bosons (fermions).
The ``$\sum$Tr'' includes a 3-momentum integral and summation over Matsubara frequencies 
(with factors of -1/$\beta $) and discrete quantum numbers.
The Luttinger-Ward functional (LWF),
\begin{equation}
\Phi=\sum_{\nu=1}^\infty \Phi_\nu\,\,,\Phi_\nu=\sum\text{Tr}\{\frac{1}{2\nu}\Sigma_\nu(G)G \ ,
\label{phi}
\end{equation}
contains the 2-body interaction contribution to the EoS. It cannot be as straightforwardly
resummed as the $ T $-matrix due to additional $1/\nu$ factors for the $\nu$-th order closed loop diagrams
represented by closed selfenergies, $\Sigma_\nu(G)$, needed to eliminate double counting.
In ladder approximation one has $ \Sigma_\nu(G)=\int d\tilde{p} \ [VG^{0}_{(2)}V G^{0}_{(2)}\cdots V]G $,
%Using the notation $\int d\tilde{p}\equiv-\beta^{-1}\sum_n\int d^{3}\textbf{p}/(2\pi )^3$ with 
%$\tilde{p}\equiv(i\omega_n,\textbf{p})$, 
so that the LWF functional $\Phi$ can be expressed as
\begin{align}
\Phi=\frac{1}{2}\sum\text{Tr} &\bigg\{G\bigg[V+\frac{1}{2}V G^{0}_{(2)}V+\ldots
+\frac{1}{\nu}VG^{0}_{(2)}V G^{0}_{(2)}\ldots .V+\ldots\bigg]G\bigg\} \ . 
\label{phi2}
\end{align}
The part in brackets, $[\cdots]$, has a structure similar to the
$T$-matrix resummation,  
\begin{eqnarray}
T&=&V+V G^{0}_{(2)}V+\ldots+VG^{0}_{(2)}V G^{0}_{(2)}\ldots V+\ldots=  
\left[\sum\limits_{\nu=0}^{\infty} \left(V G^{0}_{(2)}\right)^\nu \right] V
= [1-VG^{0}_{(2)}]^{-1}  V\ , 
\end{eqnarray}
except for the extra coefficients $1/\nu$. However, we can write  
\begin{align}
V+\frac{1}{2}V G^{0}_{(2)}V+\ldots+\frac{1}{\nu}VG^{0}_{(2)}V G^{0}_{(2)}\dots V + \dots 
=  \left[\sum\limits_{\nu=1}^{\infty}\frac{1}{\nu}\left(VG^{0}_{(2)} \right)^\nu\right] 
[G^{0}_{(2)}]^{-1} &= -\ln[1-VG^{0}_{(2)}] [G^{0}_{(2)}]^{-1}
\nonumber \\
&\equiv \text{Log}\,T
\label{logT}
\end{align}
and to obtain a matrix-log representation of the skeleton series~\cite{Liu:2016nwj,Liu:2016ysz}, as a 
generalization of usual $ T $-matrix resummation. The natural-base ``Log'' should be understood 
as a matrix operation in a  discrete energy-momentum space together with other quantum numbers,
defined through its power series~\cite{Liu:2016ysz,Liu:2016nwj}. Closing one of the external lines 
of this quantity with a $G$, in resemblance of Eq.~(\ref{eq_selfE}), we define
$\text{Log}\,\Sigma\equiv\int d\tilde{p}~\text{Log}\,T ~G$. The LWF is obtained by closing another 
external line as
\begin{align}
\Phi=\frac{1}{2}\int d\tilde{p}~G~\text{Log}\Sigma,\,\Omega
=\sum_{j}\mp d_{j} \int d\tilde{p}\Big\{\ln (-G_{j}(\tilde{p})^{-1}) 
+[\Sigma_{j}(\tilde{p})-\frac{1}{2}\text{Log}\Sigma_{j}  (\tilde{p})]G_{j}(\tilde{p})\Big\} \ , 
\end{align}
where $j$ sums over the particle species. This procedure to sum the diagrams is illustrated in 
Fig.~\ref{fig_wheel}.
The resummation of the LWF is critical to incorporate bound-state (or resonance) 
contributions to the pressure. The selfenergy of partons can then be obtained from the 
selfconsistent $T$-matrix equation via the functional derivative of this LWF, which explicitly
illustrated the  ``conserving approximation'' for the resulting 1- and 2-body spectral functions
within this scheme~\cite{Baym:1961zz,Baym:1962sx}. This, in turn, allows to probe transitions from
parton to meson degrees of freedom, if any, in the system.
%With this off-shell spectra and $ T $-matrices, the transport properties can be calculated including 
%the many-body quantum effects.
\begin{figure}[!t]
        \centering
        \includegraphics[width=1\columnwidth]{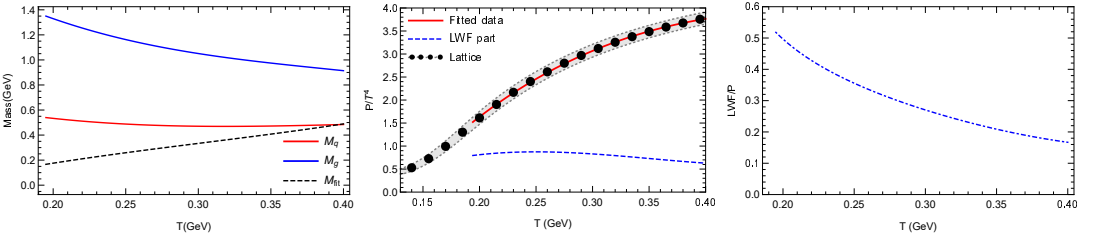}
        \caption{The fit results for the bare masses of the quarks and gluons in the Hamiltonian (left 
      panel), the resulting fit to the lQCD data~\cite{Bazavov:2014pvz} for the QGP pressure
           (middle panel; solid line: total, dashed line: LWF contribution), and the relative 
   contribution of the LWF part to the total pressure (right panel).}
        \label{fig_scs-eos}
\end{figure}

The resulting equation of state is shown in the middle panel of Fig.~\ref{fig_scs-eos} where we utilized 
the bare light-parton masses figuring in the Hamiltonian as fit parameters,
shown in the left panel in Fig.~\ref{fig_scs-eos}. 
We use the ansatz $ M_q=M_\text{fit}+M^\text{np}_q $ and $ M_g=3/2M_\text{fit}+M^\text{np}_g $ 
with a fit parameter $M_\text{fit}$ and additional mass contributions separately for quarks 
and gluons which follow from the nonperturbative part given by the infinite-distance limit of
the potential, $\tilde{V}$, in the corresponding color channel. 
This ansatz allows the masses to smoothly transit from nonperturbative behavior at low temperature 
to perturbative behavior at high temperature. At low temperatures gluons are found to essentially 
decouple from the pressure due to their large masses. On the other hand, the LWF contribution grows 
with decreasing temperature, see right panel of Fig.~\ref{fig_scs-eos}, mostly driven by the formation
of mesonic and diquark bound states / resonances. In this sense we find a gradual transition 
in the degrees of freedom from partonic to hadronic states as the temperature approaches $ T_c $ 
from above. This is different in the WCS, where the LWF does not exceed 15\% of the total pressure 
and is rather constant with temperature.

%%%%%%%%%%%%%%%%%%%%%%%%%%%%%%%%%%%%%%%%%%555
\section{Spectral properties of QGP}
\label{sec_spec}
%%%%%%%%%%%%%%%%%%%%%%%%%%%%%%%%%%%%%%%%%%%
\begin{figure}[!ht]
	\begin{center}
		\includegraphics[width=0.99\columnwidth]{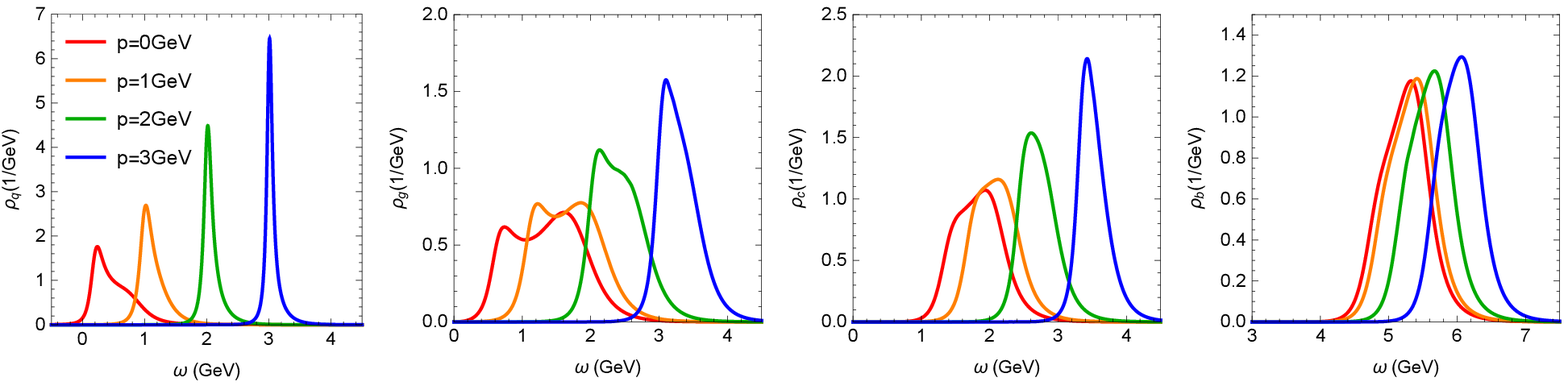}
		\includegraphics[width=0.99\columnwidth]{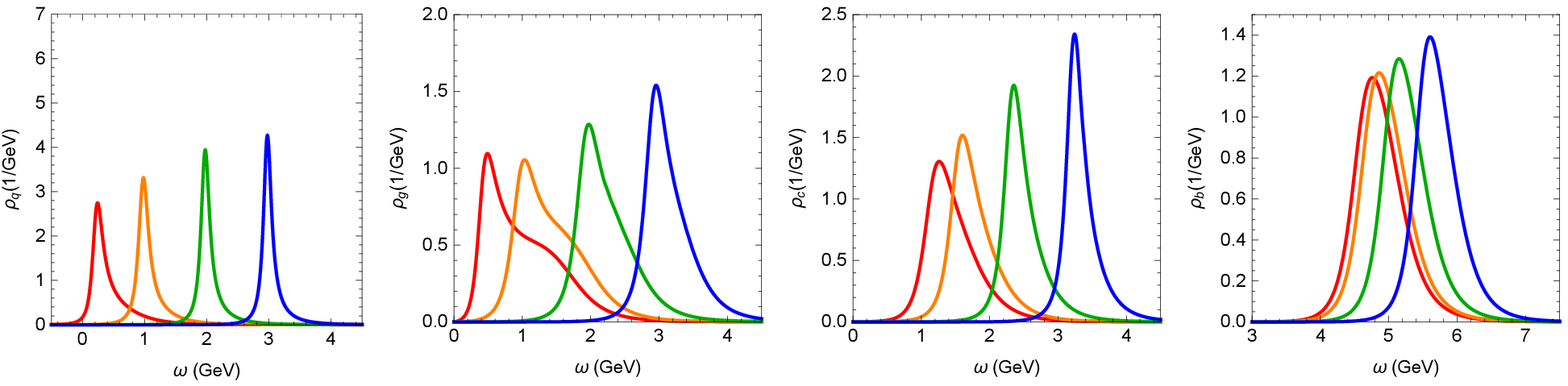}
		\caption{In-medium parton spectral functions from the selfconsistent solution in the SCS.
From left to right, we show the light-quark ($q$), gluon ($g$), charm-quark ($c$) and bottom-quark ($b$)
spectral functions for 4 different 3-momenta in each panel, and for temperatures $T$=0.194\,GeV 
(upper row) and $T$=0.400\,GeV (lower row). Note the melting of the low-momentum quark and gluon 
quasiparticles at low temperatures (with additional low-energy collective modes), while the charm 
and especially bottom quark remain good quasiparticles.} 
		\label{fig_scs-spec}
	\end{center}
\end{figure}
\begin{figure}[!b]
	\begin{center}
		\includegraphics[width=1\columnwidth]{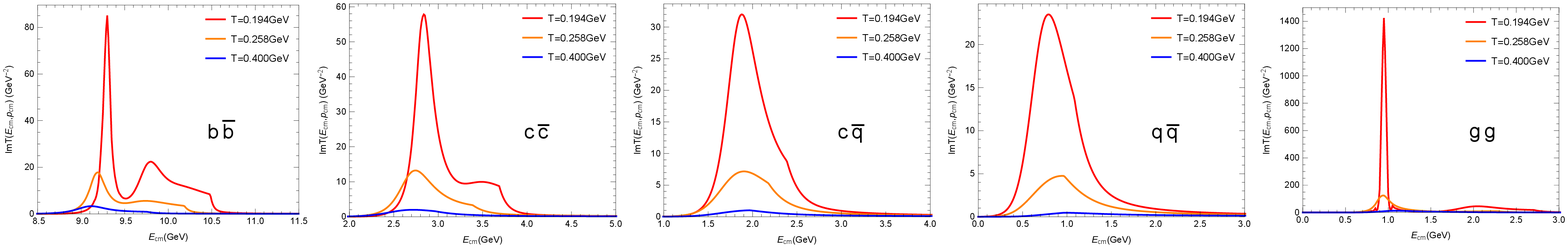}
		\caption{Imaginary part of the on-shell color-singlet $S$-wave \(T\)-matrices 
for different 3 different temperatures in each panel for: (from left to right)  
bottomonium ($b\bar b$), charmonium ($c\bar c$), $D$-meson ($c\bar q$), $\rho$-meson ($q\bar q$), 
and glueball ($gg$) channels. Note the decreasing $y$-axis scale for the first 4 panels.
%The different colors correspond to different temperatures, $T=0.194$~GeV,$T=0.258$~GeV, $T=0.400$~GeV .
		}
		\label{fig_scs-T}
	\end{center}
\end{figure}
We proceed to inspect the spectral properties underlying our selfconsistent off-shell calculations 
of the EoS discussed in the previous section, cf.~Fig.~\ref{fig_scs-spec}.
Near $T_c$, the low-momentum light-parton spectral functions acquire widths of order 1~GeV, 
significantly exceeding their masses, and thus implying a melting of their quasiparticle structure. 
Consequently, their contribution to the EoS is suppressed, while at the same time the pertinent 
$T$-matrices develop broad mesonic bound states with appreciable strength (and a mass close to 
the vacuum $\rho$ mass). These states therefore play a dual role of newly emerging degrees of 
freedom and providing large interaction strength for the partons, thus connecting the strong
coupling nature of the QGP with a gradual transition to hadronic states. Furthermore, the attractive
real part of the light-parton selfenergies leads to the appearance of low-energy collective modes 
in their spectral functions. These modes could play an important role for describing the quark 
susceptibilities where quasiparticle models face difficulties due to their large masses which 
suppress fluctuations. 
%This is a result from the nontrivial competition of the decreasing interaction strength and increase 
%of the density. 
At high parton 3-momenta and higher temperature, the quasiparticle structures re-emerge -- a 
manifestation of the weakening QCD force with increasing momentum transfer and illustrating the
changing degrees of freedom as the QGP is probed with varying resolution. 
In the WCS (not shown), all partons remain well-defined quasiparticle at all temperatures and 
3-momenta, while rather weak and sharp resonances appear in the $T$-matrices at low temperature.

\begin{figure}[!b]
	\begin{center}
		\includegraphics[width=0.49\columnwidth]{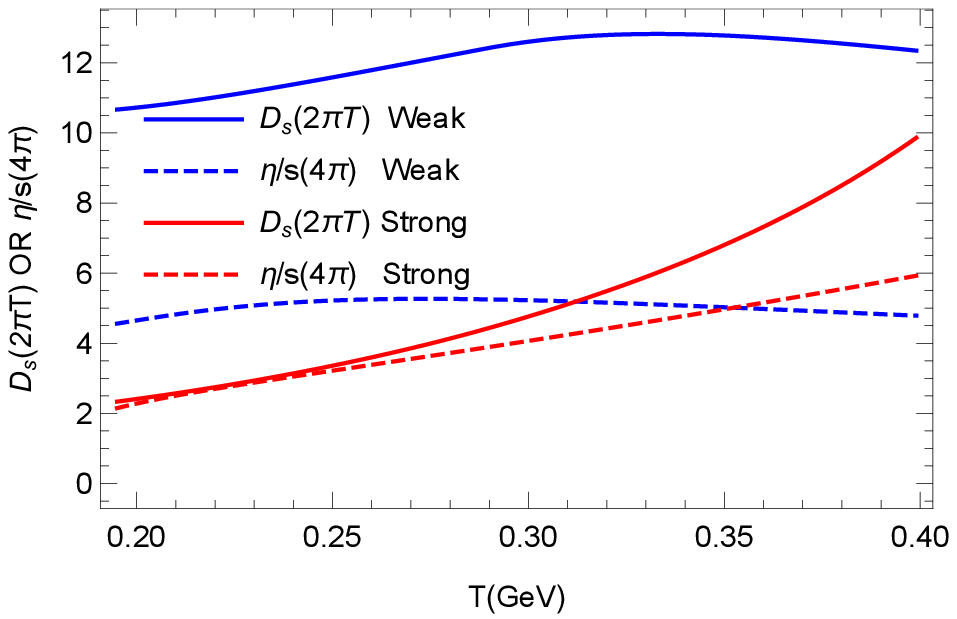}		\includegraphics[width=0.49\columnwidth]{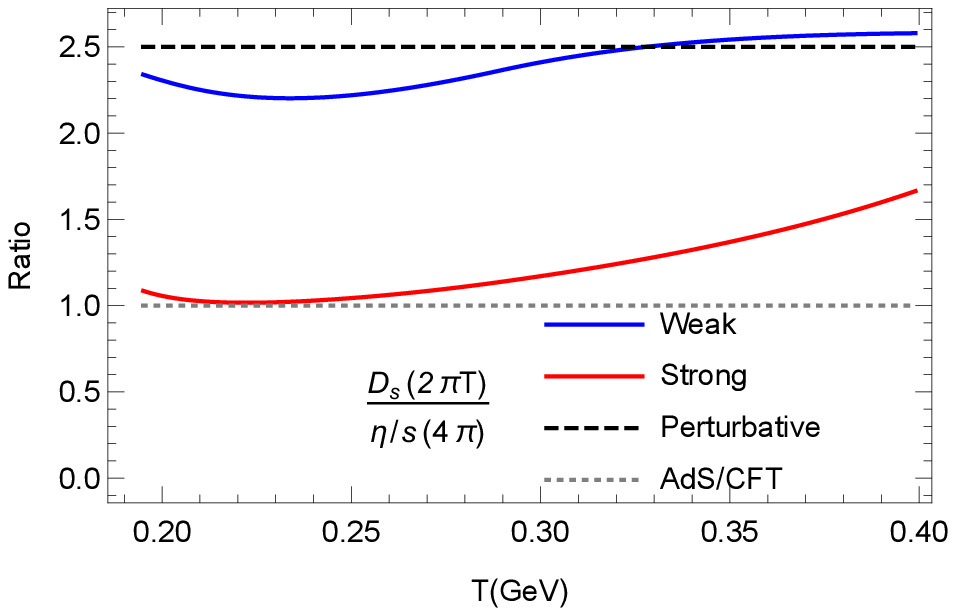}
		\caption{The transport coefficients $ D_s(2\pi T) $ and $ \eta/s(4\pi) $ of SCS and WCS (left); The ratio of $ D_s(2\pi T)$ of $ \eta/s(4\pi) $ of SCS and WCS (right).}
		\label{fig_coeff}
	\end{center}
\end{figure}

%%%%%%%%%%%%%%%%%%%%%%%%%%%%%%%%%%%%%%%
\section{Transport properties of QGP}
\label{sec_trans}
%%%%%%%%%%%%%%%%%%%%%%%%%%%%%%%%%%%%
Thus far the lQCD constraints imposed on our approach (HQ free energy, EoS and quarkonium correlator
ratios (not explicitly discussed here)) cannot decisively distinguish between the SCS and WCS. As 
another constraint we have therefore investigated transport coefficients which figure in applications
to heavy-ion phenomenology, specifically the HQ diffusion coefficient and the ratio of viscosity
over entropy density.

For the HQ friction coefficient, we extend previous $T$-matrix calculations carried out for the free and 
internal energies as potential proxies~\cite{vanHees:2007me,Riek:2010fk,Prino:2016cni}. In particular,
due to the large widths in the SCS we account for off-shell effects based on the formalism 
described in Ref.~\cite{Danielewicz:1982kk}. Schematically, one has 
\begin{align}
A(p)=&\left\langle (1-\frac{\textbf{p}\cdot\textbf{p}'}{p^2})\rho_i\rho_i\rho_c\right\rangle \ ,  
\end{align}
where $\rho_{i(c)}$ are spectral functions for light partons (charm quark) and $\textbf{p}$ 
($\textbf{p}'$) denotes the incoming (outgoing) charm-quark momentum. The spatial diffusion 
coefficient is defined as ${\cal D}_s= T/(A(0)M)$. 

For the viscosity, $\eta$, a Kubo formula is employed using the leading-density 
energy-momentum tensor~\cite{zubarev1974nonequilibrium} with relativistic 
extension, 
\begin{align}
\eta=&\lim\limits_{\omega\rightarrow 0}\sum_i\frac{\pi d_i}{\omega} \int
\frac{d^3\textbf{p}d\lambda}{(2\pi)^3} \frac{p_x^2p_y^2}{\varepsilon^2_i(p)}
\rho_i(\omega+\lambda,p)\rho_i(\lambda,p)[n_i(\lambda)-n_i(\omega+\lambda)]
\ , 
\end{align}
where $d_i$ and  $n_i(\omega)$ are the partons' degeneracies and thermal distribution functions,
respectively. The corrections from higher orders are expected to be 
small~\cite{Iwasaki:2006dr,Iwasaki:2007iv,Lang:2013lla,Ghosh:2014yea}, which we have checked to 
remain valid within our approach.

The dimensionless quantities ${\cal D}_s(2\pi T)$ and $4\pi\eta/s$ characterize the interaction 
strength of the bulk medium (with smaller values indicating stronger coupling); they are shown 
in the left panel of Fig.~\ref{fig_coeff}. For the SCS both transport coefficients are within 
a factor of two of the conjectured quantum lower bounds of one, and increase with temperature 
indicating a transition to a more weakly coupled medium. On the contrary, for the WCS both 
transport coefficients are significantly larger and rather constant with temperature. Especially 
the HQ diffusion coefficient acquires magnitudes which do not compare well with current extractions 
from HF phenomenology in high-energy heavy-ion collisions~\cite{Prino:2016cni}.

In an attempt to better quantify the notions of  "strongly" and ``weakly" coupled media, one can
inspect the ratio for the two dimensionless transport coefficients discussed above~\cite{Rapp:2009my}. 
In particular, the ratio $r\equiv  [2\pi T{\cal D}_s]/[4\pi\eta/s]$ is expected to be near 
one in the strong-coupling limit~\cite{Kovtun:2004de,Gubser:2006qh}, while perturbative
estimates~\cite{Danielewicz:1984ww} appropriate for a weakly coupled system result in $\sim$5/2. 
We plot this ratio in the right panel of Fig.~\ref{fig_coeff} for both the SCS and WCS.
Interestingly, for the SCS the ratio is around one for low temperatures, slowly increasing with
temperature but still significantly below 5/2 at $T$=400\,MeV. On the contrary, for the WCS
the ratio is close to 5/2 characteristic for a weakly-coupled system even at low temperatures,
with insignificant temperature dependence.

%\FloatBarrier
%%%%%%%%%%%%%%%%%%%%%%
\section{Conclusion}
\label{sec_concl}
%%%%%%%%%%%%%%%%%%%%%%
Unraveling the microscopic structure of the strongly coupled QCD medium in the vicinity
of the transition from hadronic to partonic matter is one of the key challenges in the study
of the QCD phase diagram. Toward this end, we have put forward a thermodynamic $T$-matrix 
formalism which includes basic nonperturbative ingredients that are expected to be relevant:
For the underlying two-body interaction we account for remnants of the confining force encoded 
in a screened Cornell potential ansatz, while the $T$-matrix allows for a ladder resummation
to account for dynamically formed bound states and retains the full off-shell properties of
one- and two-body spectral functions. Our starting point is a relativistic Hamiltonian,
whose in-medium two-body interaction is constrained by lattice-QCD data for the heavy-quark 
free energy and euclidean quarkonium correlators, while the light-parton masses in the kinetic
term are fitted to the lattice equation of state. At this level, the solution to our set-up
is not unique; however, a strongly coupled solution has the attractive features of melting
light-parton degrees of freedom in connection with emerging hadronic bound states as the 
temperature approaches $T_{\rm c}$ from above. We furthermore evaluated transport properties 
of the QGP by computing the shear-viscosity-to entropy density ratio and the heavy-quark 
diffusion coefficient. In the strongly-coupled scenario, their values are within a factor 
of $\sim$2 of the conjectured quantum lower bound, and slowly increase with temperature.
The ratio of these two quantities corroborates that the QGP near $T_{\rm c}$ is a 
strongly-coupled medium near the quantum lower bound, which reiterates the necessity of 
including quantum off-shell effects in studying its properties. Future applications include 
the calculation of high-$p_T$ parton transport, quark-number susceptibilities to explore the 
finite $\mu_q$ part of the phase diagram, and the explicit inclusion of condensate formation 
to treat chiral symmetry breaking and color superconductivity. 
\\

{\bf{Acknowledgments}}\\
This work has been supported by the U.S. National Science 
Foundation under grant no. PHY-1614484.

%%\begin{figure}[h]
%%% Use the relevant command for your figure-insertion program
%%% to insert the figure file.
%%\centering
%%\includegraphics[width=1cm,clip]{tiger}
%%\caption{Please write your figure caption here}
%%\label{fig-1}       % Give a unique label
%%\end{figure}
%
%For two-column wide figures use syntax of figure~\ref{fig-2}
%\begin{figure*}
%\centering
%% Use the relevant command for your figure-insertion program
%% to insert the figure file. See example above.
%% If not, use
%\vspace*{5cm}       % Give the correct figure height in cm
%\label{fig-2}       % Give a unique label
%\end{figure*}
%
%For figure with sidecaption legend use syntax of figure
%%\begin{figure}
%%% Use the relevant command for your figure-insertion program
%%% to insert the figure file.
%%\centering
%%\sidecaption
%%\includegraphics[width=5cm,clip]{tiger}
%%\caption{Please write your figure caption here}
%%\label{fig-3}       % Give a unique label
%%\end{figure}
%

%
% BibTeX or Biber users please use (the style is already called in the class, ensure that the "woc.bst" style is in your local directory)
% \bibliography{name or your bibliography database}
\bibliography{refcnew}
%\begin{thebibliography}{}
%%
%% and use \bibitem to create references.
%Journal Author, Journal \textbf{Volume}, page numbers (year)
%Book Author, \textit{Book title} (Publisher, place, year) page numbers
%\end{thebibliography}

\end{document}